\documentclass[11pt,oneside,letterpaper]{article}
\usepackage{amssymb}
\usepackage{amsmath}
\usepackage[dvips]{graphicx}
\usepackage{setspace}
\usepackage{fancyhdr}
\usepackage{ifpdf}
\usepackage{graphicx}
\usepackage{comment}

 \def\p{\partial}

\newcommand{\bea}{\begin{eqnarray}}
\newcommand{\eea}{\end{eqnarray}}
\newcommand{\be}{\begin{equation}}
\newcommand{\ee}{\end{equation}}

\newcommand{\bi}{\begin{itemize}}
\newcommand{\ei}{\end{itemize}}

\numberwithin{equation}{section}
\allowdisplaybreaks  % allow page breaks in displayed eqs

\begin{document}

%\vspace*{2.5cm}
%\begin{comment}
\begin{center}
{ \LARGE \textbf{Sailing from Warped AdS$_3$ to Warped dS$_3$ in Topologically Massive Gravity}\\}
%\vspace*{1.7cm}
\begin{center}
Dionysios Anninos
\end{center}
\end{center}
\vspace*{0.6cm}
\begin{center}
Jefferson Physical Laboratory, Harvard University, Cambridge, MA 02138, USA
\vspace*{0.8cm}
\end{center}
\vspace*{1.5cm}
\begin{abstract}
\noindent
{
Three-dimensional warped anti-de Sitter space in topologically massive gravity with a negative cosmological constant has been proposed to be holographically dual to a two-dimensional conformal field theory. We extend this proposal to both positive and vanishing values of the cosmological constant where stretched warped anti-de Sitter space is found to be a solution. For positive cosmological constant, another class of warped solutions is obtained by a spacelike (timelike) line fibration over Lorentzian (Euclidean) two-dimensional de Sitter space. These solutions exhibit a cosmological horizon and Hawking temperature much like de Sitter space. Global identifications of this warped de Sitter space may contain a horizon in addition to the cosmological one. At a degenerate point, warped de Sitter space becomes a fibration over two-dimensional flat space. Finally, we study scalar waves in these backgrounds. Scalars in stretched warped anti-de Sitter space exhibit superradiance which can be interpreted as Schwinger pair production of charged particles in two-dimensional anti-de Sitter space.
}
\end{abstract}
\newpage
%\onehalfspacing
%\keywords{black holes, topologically massive gravity, warped $AdS_3$, de Sitter }

\tableofcontents
\setcounter{tocdepth}{2}

\section{Introduction and Summary}

Three-dimensional gravity has proved to be a useful toy model to study questions about quantum gravity, however it suffers from the absence of local propagating degrees of freedom. The problem of no local degrees of freedom was cured by adding a gravitational Chern-Simons term to the action which leads to a consistent theory of gravity with one local degree of freedom \cite{Deser:1981wh,Deser:1982vy}. Furthermore, adding a negative cosmological constant also allows for black hole solutions \cite{Banados:1992gq,Banados:1992wn}. The most understood case is that of three-dimensional anti-de Sitter space ($AdS_3$) which has been studied extensively in the context of topologically massive gravity (TMG) \cite{Li:2008dq}.

Due to the Chern-Simons term, the number of possible vacua of the theory is far richer \cite{Nutku:1993eb,Gurses,Bouchareb:2007yx}. It was proposed in \cite{Anninos:2008fx}, for a negative cosmological constant $\Lambda < 0$, that one of these vacua known as spacelike warped-anti de Sitter space ($WAdS_3$) could have a two-dimensional holographic dual conformal field theory in analogy to the case of $AdS_3$ \cite{Brown:1986nw}. Spacelike (timelike) $WAdS_3$ is given by a spacelike (timelike) real line fibration over Lorentzian (Euclidean) $AdS_2$. The solutions are specified by two parameters: the length scale $|\Lambda|^{-1/2}$ and the warp factor $a^2$ which determines the relative scale of the fiber coordinate.

In this note, we extend the discussion of $WAdS_3$ to both a vanishing and positive cosmological constant and identify a new set of vacua for $\Lambda > 0$. The motivation for our work is to seek possible stable vacuum candidates for a greater region of parameter space in TMG and to explore de Sitter like vacua which contain cosmological horizons. We further explore global identifications of such vacua. Also, as a first step towards understanding linearized perturbations in these backgrounds, we study massive scalar waves in the various backgrounds we discuss. Linearized perturbations in the spacelike $WAdS_3$ vacuum for $\Lambda < 0$ were discussed in \cite{Anninos:2009zi}.

The first property we discuss is that both spacelike and timelike $WAdS_3$ are in fact solutions for all values of the cosmological constant. However, the warp factor can only take a particular range for each case. In particular, when $\Lambda = 0$ the warp factor can only equal four, whereas for positive (negative) $\Lambda$ it is constrained to be greater (less than) four. In fact, timelike $WAdS_3$ was first studied for $\Lambda = 0$ in \cite{Vuorio:1985ta,Percacci:1986ja} and a general analysis of non-trivial TMG solutions with $\Lambda = 0$ was given in \cite{Ortiz:1990nn}. It is worth noting that for $\Lambda < 0$ the warp factor for both spacelike and timelike $WAdS_3$ can be either \emph{stretched} ($a^2 > 1$) or \emph{squashed} ($a^2 < 1$) and $WAdS_3$ has a significantly different geometry in these two regions.

%It then follows from the analysis of \cite{Anninos:2009zi} that the spacelike stretched $WAdS_3$ vacua, with Comp\`ere-Detournay boundary conditions \cite{Compere:2009zj}, are perturbatively stable in TMG for all $\Lambda$.

Global discrete identifications of spacelike stretched warped anti-de Sitter space are studied in order to obtain the warped black hole solutions for all values of the cosmological constant \cite{Nutku:1993eb,Gurses,Bouchareb:2007yx,Moussa:2003fc}. This allows us to extend the proposal that these spacetimes are holographically dual to a conformal field theory to all possible values of the cosmological constant. We propose a new set of central charges for positive and vanishing cosmological constant and show they are consistent with the asymptotic symmetry group analysis in \cite{Compere:2008cv}. None of them depends on the intrinsic parameters of the black holes and their difference always agrees with the diffeomorphism anomaly computed holographically in \cite{Kraus:2005zm}.

We then proceed to study a qualitatively different solution given by a fibration of two-dimensional Lorentzian (Euclidean) de Sitter space over the real spacelike (timelike) line. We refer to this geometry as spacelike (timelike) warped de Sitter space ($WdS_3$). It is a solution to TMG only in the presence of a positive cosmological constant. We find the metric in both the static patch as well as the global patch. $WdS_3$ exhibits a cosmological horizon and shares various properties, such as the Hawking temperature, with the usual three-dimensional de Sitter space. The geodesics are studied and it is found that they behave qualitatively similar for all values of the warp factor, in contrast to the sharp difference in the behavior of geodesics for stretched and squashed $WAdS_3$.

Global identifications of $WdS_3$ are subsequently studied. Interestingly, we identify a class of solutions with a cosmological horizon as well as a inner horizon but no closed timelike curves (CTCs), whose metric was discovered in \cite{Nutku:1993eb,Gurses,Moussa:2003fc}. This is in contrast to the well known fact that there are no smooth asymptotically de Sitter black hole solutions in three-dimensional Einstein gravity with a positive cosmological constant. We also find that there exists a self-dual type geometry for spacelike $WdS_3$ where the fiber-coordinate is identified.

Finally, we study the scalar wave-equation in both the $WAdS_3$ and $WdS_3$ backgrounds. We find that for stretched $WAdS_3$ a superradiance-like effect occurs where the scalar acquires a non-zero incoming or outgoing flux at the boundary. No such effect is observed for the squashed case. We also obtain explicit highest weight solutions for $WAdS_3$. For the squashed or unwarped case, the conformal weight is real for all masses satisfying a Breitlohner-Freedman type bound \cite{Breitenlohner:1982bm,Breitenlohner:1982jf}. For the stretched case, the weight becomes complex for scalars with large momentum along the fiber coordinate. We interpret the violation of the Breitenlohner-Freedman bound in terms of Schwinger pair production \cite{Schwinger:1951nm} of massive charged particles in $AdS_2$ \cite{Pioline:2005pf}. The asymptotics of scalar fields in $WdS_3$ resemble those of regular de Sitter space with an oscillatory decaying time dependence near the boundary. The analogous conformal weight becomes complex for large values of the mass or momentum along the fiber coordinate which immediately brings the discussions of scalar fields in de Sitter space to mind \cite{Strominger:2001pn}.

Below we present a list of some of the known TMG vacua for positive, negative and vanishing cosmological constants. Both timelike and spacelike $WAdS_3$ are solutions for all values of $\Lambda$ whereas timelike and spacelike $WdS_3$ are only solutions for $\Lambda > 0$ and null warped $AdS_3$ is only a solution for $\Lambda < 0$. The usual $dS_3$, $AdS_3$ and flat space solutions are the only vacua shared by both Einstein gravity and TMG.
\begin{eqnarray}
\nonumber \Lambda < 0 &:& \{ AdS_3, WAdS_3, Null_\pm \} \\
\nonumber \Lambda = 0 &:& \{ \mathbb{R}^{2,1}, WAdS_3 \}\\
\nonumber \Lambda > 0 &:& \{ dS_3, WAdS_3, WdS_3, W\mathbb{R}^{2,1} \}
\end{eqnarray}
%\begin{figure}[h]
%\begin{center}
% \input{RNMeQ.tpx}
% \end{center}
% \caption{Phase Diagram of the Warped Vacua of TMG.}
%\end{figure}
\begin{figure}[h]\label{fig}
  \begin{center}
    \includegraphics[height=3.0in]{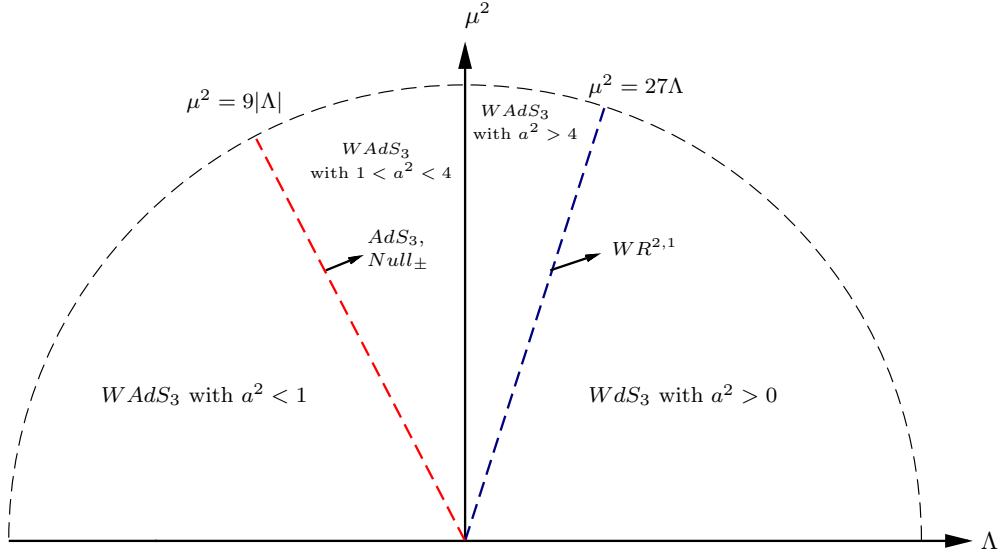}
  \end{center}
  \caption{\small Sea of Warped Vacua in the TMG Parameter Space. }
  \label{fig-label}
\end{figure}

We should note that $WAdS_3$ is a solution of TMG for $\Lambda = +1/\ell^2$ only in the region $\mu^2 \ell^2 > 27$,
where $\mu$ is the Chern-Simons coefficient, and $WdS_3$ is only a solution in the parameter region $\mu^2 \ell^2 < 27$.
The point $\mu^2 \ell^2 = 27$ is special in the sense that the warp factor diverges and neither $WAdS_3$ or $WdS_3$ seem to
be a solution at that point. At this point, we can take a rescaling limit to find that the solution becomes a fibration over
two-dimensional Minkowski space. We refer to this solution as warped flat space ($W\mathbb{R}^{2,1}$). On the other hand, $WAdS_3$ is a solution for all $\Lambda < 0$ and reduces to $AdS_3$ at $\mu^2\ell^2 = 9$. At the chiral point $\mu^2\ell^2 = 1$ the $WAdS_3$ solution has $a^2 < 1$. These properties are shown in figure \ref{fig}.

\section{Framework and Geometry}

Our story begins with the action of topologically massive gravity (TMG) \cite{Deser:1981wh,Deser:1982vy} which is given by
\be
I_{TMG} = \frac{1}{16\pi G}\int d^3x \sqrt{-g}\left[  R - 2\Lambda + \frac{1}{2\mu}\varepsilon^{\lambda \mu \nu} \Gamma^\rho_{\lambda \sigma} ( \partial_\mu \Gamma^\sigma_{\rho\nu} + \frac{2}{3} \Gamma^\sigma_{\mu\tau} \Gamma^{\tau}_{\nu \rho} ) \right]
\ee
The non-linear equations of motion for TMG with a cosmological constant $\Lambda = \pm 1/\ell^2$ or $\Lambda = 0$ are given by
\be
R_{\mu\nu} - \frac{1}{2}R g_{\mu\nu} + \Lambda g_{\mu\nu} + \frac{1}{\mu}C_{\mu\nu} \equiv \mathcal{G}_{\mu\nu} + \frac{1}{\mu}C_{\mu\nu} = 0
\ee
where
\be
C_{\mu \nu} = {{\varepsilon_{\mu}}^{\alpha \beta}}\nabla_{\alpha}(R_{\beta\nu} - \frac{1}{4}g_{\beta\nu}R)
\ee
is the Cotton tensor and $\varepsilon^{\mu\nu\rho}$ is the Levi-Civita tensor. We take $\nu \equiv \mu \ell / 3$ to be positive in what follows.

\subsection{Geometry}

The basic geometry we will focus on is the vacuum solution known as spacelike warped anti-de Sitter space, with metric:
\be
ds^2 = L^2 \left[ -(1+r^2)d\tau^2 + \frac{dr^2}{1+r^2} + a^2 (du + r d\tau)^2 \right]\label{genmetric}
\ee
where $(r,\tau,u) \in[-\infty,\infty]$. The boundary resides at $r=\pm \infty$ with fixed $u$ and also at $u=\pm\infty$ for fixed $r$. The parameter $a^2$ is known as the warp factor and can in principle take the value $a^2 \ge 0$. When the warp factor is greater (smaller) than unity the spacetime is known as stretched (squashed) spacelike $WAdS_3$. For $a^2 = 1$ the geometry reduces to $AdS_3$ in global coordinates. The Ricci scalar of the above geometry is constant and given by
\be
R = \frac{a^2 - 4}{2 L^2}
\ee
Thus, the solution has negative curvature only in the region $a^2 < 4$, positive curvature in the region $a^2 > 4$ and vanishing curvature for the region $a^2 = 4$.

We further note that for the particular geometry there is an $SL(2,\mathbb{R})_{R}\times U(1)_{L}$ isometry where the $U(1)_L$ is non-compact. The Killing vectors are given by \cite{Bengtsson:2005zj}
\begin{eqnarray}
\tilde{J}_1 &=& 2\sin \tau \frac{r}{\sqrt{1+r^2}} \partial_\tau - 2\cos\tau \sqrt{1+r^2} \partial_r + \frac{2\sin\tau}{\sqrt{1+r^2}}\partial_u\\
\tilde{J}_2 &=& -2\cos \tau \frac{r}{\sqrt{1+r^2}} \partial_\tau - 2\sin\tau \sqrt{1+r^2} \partial_r - \frac{2\cos\tau}{\sqrt{1+r^2}}\partial_u\\
\tilde{J}_0 &=& 2 \partial_\tau; \quad J_2 = 2 \partial_u
\end{eqnarray}
where the quantities with a tilde generate the $SL(2,\mathbb{R})_R$ and $J_2$ generates the $U(1)_L$.

%For large $r$ the metric asymptotes to
%\be
%ds^2 = L^2\left[ (a^2 - 1)r^2  d\tau^2 + \frac{dr^2}{r^2} + 2 a^2 r du d\tau + a^2 du^2 \right]
%\ee

The metric \ref{genmetric} is geodesically complete \cite{Rooman:1998xf,Bengtsson:2005zj} and furthermore only the constant $\tau$ slices are spacelike for all $r$ thus rendering $\tau$ as our global time coordinate. When $a^2 \leq 1$, $\partial_\tau$ is a globally timelike Killing vector, however for $a^2 > 1$ it becomes spacelike for large $r$. Thus our spacetime for $a^2 > 1$ looks like the region within the ergosphere of a rotating black hole and there can be no static observers. It is global identifications of this spacetime with $a^2 > 1$ that give rise to black hole solutions with no CTCs outside the horizon \cite{Anninos:2008fx}.

Finally, we would like to point out that the geodesics of spacelike $WAdS_3$ behave qualitatively different for $a^2$ greater or less than one. For $a^2 > 1$ the null geodesics all escape to infinity as well as some of the timelike geodesics. For $a^2 < 1$ null geodesics with non-zero momentum along $u$ are confined as well as all the timelike geodesics.

\section{Warped anti-de Sitter Space}

As was already mentioned in the introduction, $WAdS_3$ appears as a vacuum solution to TMG for all values of the cosmological constant. In terms of the TMG parameters the metric is found to be
\be
ds^2 = \frac{\ell^2}{(\nu^2 + 3\epsilon)}\left[ -d\tau^2(1+r^2) + \frac{dr^2}{1+r^2} + \frac{4\nu^2}{\nu^2 + 3\epsilon}\left( du + r d\tau \right)^2 \right]\label{wads}
\ee
where $\epsilon = \{-1,0,+1\}$ for $\Lambda = -\epsilon/\ell^2$ and $L^2 = \ell^2/(\nu^2+3\epsilon)$. The Ricci scalar of this metric is $R = -6\epsilon/\ell^2$ and thus shares the sign of $\Lambda$ as implied by the TMG equations of motion. Notice the following:
\begin{itemize}
\item When $\Lambda < 0$ the warp factor $a^2 = 4\nu^2/(\nu^2+3)$ satisfies the bound $a^2 < 4$. The special point in the parameter space is clearly $\nu^2=1$ where the metric reduces to $AdS_3$.

\item When $\Lambda > 0$ the warp factor $a^2 = 4\nu^2/(\nu^2-3)$ satisfies the bound $a^2 > 4$ for $\nu^2 > 3$. When $\nu^2 \leq 3$ the geometry has a different interpretation which is discussed in sections \ref{wfs} and \ref{wds}.

\item When $\Lambda = 0$ the warp factor $a^2 = 4$ is fixed and lies between the two aforementioned bounds.

\end{itemize}
We can also obtain the timelike warped vacua by analytically continuing $\tau \to i u$ and $u \to -i\tau$ in \ref{wads}. However, timelike $WAdS_3$ is free of CTCs only when $a^2 < 1$ \cite{Rooman:1998xf}, and this can only occur for $\Lambda < 0$. We will now continue on to discuss the warped black hole solutions for all values of $\Lambda$.

\subsection{Warped Black Holes}

Discrete global identifications of $WAdS_3$ along Killing directions were shown \cite{Anninos:2008fx} to give the warped black hole solution \cite{Nutku:1993eb,Gurses,Bouchareb:2007yx,Moussa:2003fc} which for general $\Lambda = -\epsilon/\ell^2$ become
\begin{multline}
\frac{ds^2}{\ell^2}=dt^2+\frac{ dr^2}{(\nu^2+3\epsilon)(r-r_{+})(r-r_{-})} - \left(2\nu r  - \sqrt{r_{+}r_{-}(\nu^2+3\epsilon)}\right)dtd\theta \\
+\frac{r}{4}\left(3(\nu^2 - \epsilon)r+(\nu^2+3\epsilon)(r_+ + r_-) - 4\nu\sqrt{r_+r_-(\nu^2+3\epsilon)}\right)d\theta^2
\label{eq:ng}
\end{multline}
with $r > 0$\footnote{We can in fact extend $r$ to negative values, to get the geodesically complete spacetime, see for instance \cite{Moussa:2008sj}.}, $t \in [-\infty,\infty]$ and $\theta \sim \theta + 2\pi$. The inner and outer horizons are given by $r_-$ and $r_+$ respectively. As for the vacuum geometry, for $\Lambda>0$ we consider these solutions in the region $\nu^2>3$.

The Killing direction identified is the following
\be
\partial_\theta = \pi \ell \left( T_L J_2 - T_R \tilde{J}_2 \right)\label{killdir}
\ee
where
\be
T_R = \frac{(\nu^2 + 3\epsilon)(r_+ - r_-)}{8\pi \ell}, \quad T_L = \frac{(\nu^2 + 3\epsilon)}{{8\pi \ell}}\left(r_+ + r_- -\frac{\sqrt{r_-r_+(\nu^2+3\epsilon)}}{\nu}\right)
\ee
are interpreted as the left and right moving temperatures of the thermal state that the black holes correspond to in the proposed dual CFT \cite{Anninos:2008fx}.

The Chern-Simons corrected entropy \cite{Kraus:2005vz,Solodukhin:2005ah,Park:2006gt,Tachikawa:2006sz} of the warped black hole, computed in \cite{Bouchareb:2007yx}, is then given suggestively by
\be
S_{TMG} = \frac{\pi^2 \ell }{3}\left(  T_L c_L + T_R c_R \right)
\ee
where
\be
c_R = \frac{(5\nu^2+3\epsilon)\ell}{G\nu (\nu^2+3\epsilon)}; \quad c_L = \frac{4\nu \ell}{G(\nu^2 + 3\epsilon)}\label{ccwads}.
\ee
This is precisely the form of the Cardy formula for a 2d CFT with right and left moving central charges $c_L$ and $c_R$.

As simple and heuristic consistency checks, \cite{Anninos:2008fx} noted that the central charges do not depend on the intrinsic properties of the black hole,  and their difference matched the diffeomorphism anomaly computed holographically by \cite{Kraus:2005zm}
\be
c_R - c_L = \frac{\ell}{G \nu}\label{diffan1}
\ee
which is \emph{independent} of $\epsilon$. For $\Lambda < 0$, they reduce to the known central charges of $AdS_3$ for $\nu^2 \to 1$. Furthermore, for $\nu \to \infty$ we find that both central charges tend to zero which is consistent with the fact that this is the limit where gravity becomes highly curved.

These observations suggest that the theory is holographically dual to a two-dimensional CFT with the given central charges \cite{Anninos:2008fx}. By studying the asymptotic symmetries, \cite{Compere:2008cv} have managed to verify $c_R$ for a particular set of boundary conditions. Clearly, these observations and the analysis of the asymptotic symmetries carry forward to all values of the cosmological constant.

%It will be useful to express the above central charges in terms of the warp factor $a$ and the length scale $L$ given in \ref{genmetric}
%\be
%c_R = \frac{2(1+a^2)L}{a G}; \quad c_L = \frac{2 a L}{G}\label{gencc}
%\ee

\subsection{Einstein Central Charges}

One can also compute the Bekenstein-Hawking entropy in the case that the above metric appears as a solution to theories of gravity coupled to matter but with no gravitational Chern-Simons term \cite{Anninos:2008qb}. We find that it satisfies the following equation
\be
S_{BH} = \frac{\pi^2\ell}{3}\left( \tilde{c}_L T_L +\tilde{c}_R T_R \right)
\ee
with
\be
\tilde{c}_L = \tilde{c}_R = \frac{3 a L}{G}\label{bhcc} %\frac{6\nu\ell}{G(\nu^2+3)}
\ee
We can propose these as the relevant central charges for theories of gravity coupled to matter but no gravitational Chern-Simons term. Notice that when $a^2 = 1$ they agree with the well known Brown-Henneaux central charges \cite{Brown:1986nw} since $L^2 = \ell^2/4$ in that case. The right moving central charge was verified in \cite{Compere:2007in} for the case of topologically massive electrodynamics in the region $a^2 < 1$\footnote{We should note that for $a^2 < 1$ the black holes have naked CTCs and it is not clear whether they are part of the physical spectrum. Thus the proposal is much weaker for $a^2 < 1$.}. Note that they are in fact equal due to the absence of the diffeomorphism anomaly.

\section{Warped Flat Space}\label{wfs}

If we look at the warp factor for $\Lambda>0$, we notice that $\nu^2 = 3$ is a special point where it diverges. In this brief section, we proceed to study the solution that appears for $\nu^2 = 3$ by taking a rescaling limit. Furthermore, we consider identifications of the resulting vacuum and find solutions with horizons.

By redefining the coordinates in \ref{wads} as: $r \to \sqrt{3-\nu^2}\tilde{r}$, $\tau \to \sqrt{3-\nu^2} \tilde{\tau}$ and $u \to (3-\nu^2)\tilde{u}$ and taking the limit $(3-\nu^2) \to 0$ we find that \ref{wads} becomes
\be
ds^2 = \ell^2 \left[ -d\tilde{\tau}^2 + d\tilde{r}^2 + 12 \left( d\tilde{u} + \tilde{r} d\tilde{\tau} \right)^2 \right]\label{wflat}
\ee
where $\tilde{\tau},\tilde{r},\tilde{u} \in [-\infty,\infty]$ and the Ricci scalar of the above metric is given by $R = 6/\ell^2$. We can describe the geometry of \ref{wflat}, which is a solution to TMG with $\Lambda > 0$ only when $\nu^2 = 3$ (i.e. $\mu^2\ell^2 = 27$), as a fibration over two-dimensional flat space along a fiber coordinate spanning the real line. We refer to this solution as warped flat space or $W\mathbb{R}^{2,1}$. Global identifications of $W\mathbb{R}^{2,1}$ were discussed in \cite{Moussa:2008sj} in the context of TMG coupled to a Maxwell theory with an electromagnetic Chern-Simons term. Indeed, following \cite{Moussa:2008sj}, we find that
\be
ds^2 = \ell^2\left[ - 24\xi \rho  dt^2 + \frac{1}{24\xi}\frac{d\rho^2}{\rho} + 12 \left( du + (\rho + \omega + \xi)dt \right)^2 \right]
\ee
is related to \ref{wflat} by a coordinate transformation. The above solution has a Killing horizon at $\rho = 0$.  If we identify $t \sim t + 2\pi$ the solution becomes a causally regular black hole in the parameter range $\omega > -\xi/2$, where CTCs are absent.

As usual we also find that the `timelike' case,
\be
ds^2 = \ell^2 \left[ d\tilde{x}^2 + d\tilde{y}^2 - 12 \left( d\tilde{u} - \tilde{x} d\tilde{y} \right)^2 \right]
\ee
is a solution for $\Lambda > 0$ and $\nu^2 = 3$. The `timelike' solution was found in \cite{Nutku:1993eb} where the special point $\nu^2 = 3$ was also noticed. Furthermore, \cite{Nutku:1993eb} obtained the `timelike' solution in global coordinates:
\be
ds^2 = \ell^2 \left[ dr^2 + r^2 d\theta^2 - 3 (dt + r^2 d\theta)^2 \right]
\ee
Notice that we must impose the periodic identification $\theta \sim \theta + 2\pi$ for regularity at the origin. The periodic identification of $\theta$ gives rise to CTCs for large values of $r$.

\section{Warped de Sitter Space}\label{wds}

We now move on to study a non-Einstein de Sitter like solution, i.e. with a cosmological horizon, and discuss some of its properties. Interestingly, we will find that global identifications may lead to smooth, causally regular solutions containing a horizon in addition to the cosmological one. It is worth noting that this does not occur for global identifications of three-dimensional de Sitter space which contains \textit{no} smooth BTZ like solutions\footnote{Very interesting asymptotically $dS_3$ black hole solutions have been found in new massive gravity \cite{Bergshoeff:2009hq,Bergshoeff:2009aq,Oliva:2009ip}}.

When we have a positive cosmological constant and $\nu^2 < 3$, the $r$ coordinate in \ref{wads} becomes timelike and the geometry has a significantly different interpretation. In fact, the base space that is fibered over is two-dimensional de Sitter space, $dS_2$. To go to the static patch, we can analytically continue to $r\to i r$ and $\tau \to - i t$ to obtain
\be
ds^2 = \frac{\ell^2}{(3-\nu^2)}\left[ -dt^2(1-r^2) + \frac{dr^2}{(1-r^2)} + \frac{4\nu^2}{(3-\nu^2)}\left( du + r dt \right)^2 \right]\label{wdsmetric}
\ee
with $r,t,u \in[-\infty, \infty]$. This is an everywhere Lorentzian metric which is precisely a real line fibration over two-dimensional de Sitter space. We can rewrite the warped de Sitter metric as
\be
{ds^2} = L^2 \left[ -{dt^2}(1-r^2) + \frac{dr^2}{(1-r^2)} + a^2(du+r dt)^2 \right]
\ee
which has a Ricci scalar $R = (4+ a^2)/2 L^2$ which is always positive. Thus, this solution can only be supported in TMG by a positive cosmological constant. This is unlike warped anti-de Sitter space which, as we saw, can be supported by a positive, negative or vanishing cosmological constant.

Another difference with warped anti-de Sitter space is that the above metric does not reduce to $dS_3$ for any value of $\nu$. On the other hand, when $r^2 = 1$ the metric exhibits a cosmological horizon as is usual for de Sitter space. In subsection \ref{wdsglob} we will also write the metric in global coordinates.

The isometries of the above metric (for $r^2 < 1$) are given by
\begin{eqnarray}
\tilde{J}_1 &=& -2\cosh t \frac{r}{\sqrt{1-r^2}} \partial_t - 2 \sinh t \sqrt{1-r^2} \partial_r + \frac{2 \cosh t}{\sqrt{1-r^2}} \partial_u \\
\tilde{J}_2 &=& -2\sinh t \frac{r}{\sqrt{1-r^2}} \partial_t - 2 \cosh t \sqrt{1-r^2} \partial_r + \frac{2 \sinh t}{\sqrt{1-r^2}} \partial_u \\
\tilde{J}_0 &=& 2\partial_t; \quad  J_2 = 2\partial_u
\end{eqnarray}
As in the case of $WAdS_3$ they comprise of an $SL(2,\mathbb{R})\times U(1)$ due to the fact that the isometries of $dS_2$, which are given by an $SL(2,\mathbb{R})$, are preserved by the fibration.

\subsection{Identifications of $WdS_3$}

As before, we can consider discrete identifications of the geometry along Killing directions. We immediately note that there will exist self
dual type solutions upon identifying $u \sim u + 2\pi\alpha$ \cite{Coussaert:1994tu} for both timelike and spacelike $WdS_3$.
%For timelike $WdS_3$ we will see that such an identification gives rise to CTCs.

Identifying along more general Killing directions would give rise to particle like objects with conical defects reminiscent to those studied in \cite{Park:1998qk, Park:2007yq, Balasubramanian:2002zh,Balasubramanian:2001nb} or objects with CTCs. Indeed, the warped black hole metric for $\nu^2 < 3$ becomes
\begin{multline}
\frac{ds^2}{\ell^2}=dt^2+\frac{ dr^2}{(\nu^2-3)(r-r_{+})(r+r_{-})} - \left(2\nu r  + \sqrt{r_{+}r_{-}(-\nu^2+3)}\right)dtd\theta \\
+\frac{r}{4}\left(3(\nu^2 + 1)r+(\nu^2-3)(r_+ - r_-) + 4\nu\sqrt{r_+r_-(-\nu^2+3)}\right)d\theta^2\label{wdsbh}
\end{multline}
where $\theta \sim \theta + 2\pi$. The above metric has the same coordinate invariants as $WdS_3$. Note that for the above metric to be real in the region $\nu^2 <  3$ we require $r_+$ and $r_-$ to be both positive (or both negative). Since $g_{\theta\theta}$ has two real roots, $r_{s1}=0$ and $r_{s2}$, the above metric contains CTCs in the region $r \in (r_{s1},r_{s2})$ which are not shielded by the inner horizon at $r = -r_-$.

More interestingly, there is also a solution of TMG with $\Lambda > 0$ that contains horizons, but no CTCs for certain regions of parameter space. Its metric is given by \cite{Bouchareb:2007yx}
\begin{multline}
\frac{ds^2}{\ell^2} = dt^2 + \frac{d\tilde{r}^2}{(\nu^2-3)(\tilde{r}-r_h)(\tilde{r}+r_h)}- \left(2\nu  \tilde{r} + \frac{3(\nu^2+1)\omega}{2\nu} \right) dt d\theta \\ + \frac{3(\nu^2 + 1)}{4} \left( \tilde{r}^2 + 2 \tilde{r} \omega + \frac{(\nu^2-3)r_h^2}{3(\nu^2+1)}  + \frac{3(\nu^2 + 1)\omega^2}{4\nu^2} \right)d\theta^2\label{wdsbc}
\end{multline}
where $\theta \sim \theta + 2\pi$ and $r_h > 0$. The above metric is related to \ref{wdsbh} by a simple coordinate transformation for the region $\omega^2 < 4\nu^2 r_h^2/{3(\nu^2+1)}$ (see appendix \ref{appbcwds}). Notice that even though there is a cosmological horizon at $\tilde{r} =  r_h$ and an additional horizon (with a different Hawking temperature) at $\tilde{r} = - r_h$, there are no CTCs in the parameter region
\be
\omega^2 > \frac{4 \nu^2 r_h^2}{3(\nu^2 + 1)}
\ee
thus rendering the solutions non-pathological in this region of parameter space. Both \ref{wdsbh} and \ref{wdsbc} share the same asymptotic structure. The thermodynamic properties of \ref{wdsbc} for $\nu^2 > 3$ were explored in \cite{Bouchareb:2007yx}, and we hope to study the thermodynamics for $\nu^2 < 3$ in a future work. It would also be interesting to explore the asymptotic symmetry group of $WdS_3$ and see whether the entropy of the above solutions can be reproduced by a Cardy formula.

\subsection{Global Coordinates}\label{wdsglob}

We can write the metric \ref{wdsmetric} in global coordinates as follows,
\be
ds^2 = \frac{\ell^2}{(3-\nu^2)}\left[-{d\tau^2} + \cosh^2\tau d\phi^2 + \frac{4\nu^2}{(3-\nu^2)}(du-\sinh\tau d\phi)^2 \right]
\label{wdsglobal}
\ee
where $\phi \sim \phi + 2\pi$. Note that asymptotically, for large $\tau$ the boundary is given by spacelike circles at $\tau \to \pm \infty$ for all $\nu^2 < 3$. As $\nu \to 0$ we the metric degenerates to $dS_2\times \mathbb{R}$. The isometries of the above metric are given by
\begin{eqnarray}
\tilde{J}_1 &=& 2\cos \phi  \partial_\tau - 2 \sin \phi \tanh \tau \partial_\phi + \frac{2 \sin \phi}{\cosh \tau} \partial_u \\
\tilde{J}_2 &=& 2\sin \phi  \partial_\tau + 2 \cos \phi \tanh \tau \partial_\phi - \frac{2 \cos \phi}{\cosh \tau} \partial_u \\
\tilde{J}_0 &=& 2\partial_\phi; \quad J_2 = 2\partial_u
\end{eqnarray}

If we analytically continue \ref{wdsglobal} by taking $\tau \to i\theta$ and $u \to i u$ we get the following metric:
\be
ds^2 = \frac{\ell^2}{(3-\nu^2)}\left[  d\theta^2 + \cos^2 \theta d\phi^2 - \frac{4\nu^2}{(3-\nu^2)}\left(  du + \sin\theta d\phi \right)^2 \right]
\ee
This metric can be referred to as a timelike warped de Sitter space and it is also a solution to TMG with $\Lambda > 0$. It amounts to a real line fibration over the two-sphere, i.e. Euclidean $dS_2$. For $\nu^2 \to 0$ this metric degenerates to $S^2 \times \mathbb{R}_{time}$. Note that if $\phi$ is identified, timelike $WdS_3$ has CTCs for $\nu^2 \in (0,3]$.

\subsection{Geodesics}

We would like to study the null geodesics of global warped de Sitter space. There are two conserved quantities given by the two manifest isometries $\xi^\mu_{(\phi)} = \partial_\phi$ and $\xi^\mu_{(u)} = \partial_u$
\be
g_{\mu\nu} \xi^{\mu}_{(\phi)}\frac{d}{d \lambda} \tau^\nu(\lambda) = E; \quad  g_{\mu\nu} \xi^{\mu}_{(u)}\frac{d}{d \lambda} \tau^\nu(\lambda) = p
\ee
The equation governing the null geodesics is given by
\be
p^2+a^2 (E \text{sech} \tau +p \tanh \tau )^2-a^2 L^4 \left(\frac{d\tau}{\lambda}\right)^2 = 0
\ee
where $a^2 = 4\nu^2/(3-\nu^2)$ and $L^2 = \ell^2/(3-\nu^2)$. Finding explicit solutions for $\tau(\lambda)$ is not an easy task. However, we can solve the equation asymptotically and find
\be
\tau(\lambda) \sim \pm p \sqrt{\frac{1+a^2}{a^2}}\lambda
\ee
It is also easy to see that $u(\lambda) \to \infty$ only for infinite $\lambda$ and thus our spacetime is geodesically complete. Note that the geodesics behave qualitatively the same for all $a^2$ unlike the case of $WAdS_3$.

\section{Warped de Sitter Thermodynamics}

Having discussed the $WdS_3$ metric, we are in a position to explore the cosmological horizon and its associated Hawking temperature. We also consider a particular Euclideanization to the solution, leading to the squashed three-sphere, that solves a set of real TMG equations of motion.

We begin by noting some further properties of the metric in the static patch coordinate system. There are two manifest Killing vectors, $\partial_t$ and $\partial_u$. The norm of $\partial_t$ is given by
\be
|\partial_t|^2 = -(1-r^2(a^2+1))
\ee
Thus when $r^2 < 1/(a^2+1)$, $\partial_t$ has negative norm and at $r^2=1/(a^2+1)$ we find its norm is vanishing. Thus we identify $r^2=1/(a^2+1)$ as a stationary limit surface. Constant $r$ hypersurfaces are found to be timelike for $r^2<1$ and spacelike for $r^2>1$. At $r^2=1$, $g^{rr}=0$ and there is a cosmological horizon for the static patch observer. %At the horizon, $\partial_t$ is spacelike.

The vector $\chi^\mu = \xi_{(t)}^\mu + \omega \xi_{(u)}^\mu$ has vanishing norm at the horizon for $\omega = -1$. From $\chi^\mu$ we can obtain the surface gravity at the cosmological horizon
\be
\kappa^2 = -\frac{1}{2}(\nabla_\mu \chi_\nu)(\nabla^\mu \chi^\nu) = 1
\ee
Having obtained the surface gravity at the horizon we can read off the Hawking temperature which turns out to be
\be
T_{WdS} = \frac{1}{2\pi L} = \frac{\sqrt{3 - \nu^2}}{2 \pi \ell}
\ee

The existence of a cosmological horizon at $r^2 = 1$ in the static patch coordinates implies an associated entropy. The entropy will consist of the usual Bekenstein-Hawking piece as well as a correction due to the gravitational Chern-Simons term. We can compute the area of the horizon at fixed $r$ and $t$ and find that it diverges as $\int du$. Thus, the entropy itself diverges in the same fashion. This is qualitatively different from the $dS_3$ case whose entropy is finite.

In the case of the self dual solution, with $u \sim u + 2\pi$ we can compute the Chern-Simons corrected entropy. For $\nu^2 < 3$, we find it to be
\be
S_{WdS} = \frac{2 \pi \nu \ell}{3 G  (3-\nu^2)}
\ee
It is tempting to relate the periodicity of $u$ to a left moving temperature. Using the same prescription as in \ref{killdir}, we find $T_L = 1/2\pi\ell$ such that the entropy reads
\be
S_{WdS} = \frac{\pi^2 \ell}{3} c_L T_L ; \quad c_L \equiv \frac{4\nu \ell}{G(3-\nu^2)}
\ee
This is the form of the entropy of two-dimensional conformal field theory with left moving central charge $c_L$ at temperatures $T_L$ and $T_R = 0$. We hope to explore this suggestive observation in future work.

\subsection{Euclideanization}

One can also consider a particular Euclideanization of $WdS_3$ by taking $\tau \to i\theta$, $u \to -i \psi$ and $\nu \to \tilde{\nu} \equiv i\nu$ leading to the metric of the squashed three-sphere:
\be
ds^2 = \frac{\ell^2}{(3+\tilde{\nu}^2)}\left[  d\theta^2 + \cos^2 \theta d\phi^2 + \frac{4\tilde{\nu}^2}{(3+\tilde{\nu}^2)}\left(  d\psi - \sin\theta d\phi \right)^2 \right]
\ee
When $\tilde{\nu}^2 = 1$ the metric becomes that of the three-sphere expressed as a fibration which is the Euclideanization of $dS_3$. The identifications of the squashed sphere are generated by $(\theta, \psi) \sim (\theta+2\pi,\psi \pm 2\pi)$. One can see this by looking at the metric near $\theta \sim -\pi/2$ and $\theta \sim +\pi/2$  and considering loops in the $\theta$ direction keeping $\psi \pm \phi$ fixed\footnote{We thank Frederik Denef for clarifying this point.}. We can also consider lens spaces of the above by changing the identifications of $\psi$.

Note that the above metric is a solution to the following Euclidean TMG equations of motion
\be
% + \frac{i}{\mu } C_{ij} = \mathcal{G}_{ij} - \frac{\ell}{3\tilde{\nu}} C_{ij} = 0 \mathcal{G}_{ij}
\left(R_{ij} - \frac{1}{2}R g_{ij} + \frac{1}{\ell^2}g_{ij}\right) + \frac{1}{\tilde{\mu}} C_{ij} = 0
\ee
where we have defined $\tilde{\mu} \equiv  3\tilde{\nu}/\ell$. The Euclidean equations of motion above are real and thus somewhat different from the typical Euclidean equations of motion coming from Chern-Simons theories which are complex.

It is interesting to compare the above Euclideanization of $WdS_3$ to one which is perhaps more relevant to $WAdS_3$. If we analytically continue $WAdS_3$ for $\Lambda < 0$ by taking $\tau \to i \tau$, $u \to -i u$ and $\nu \to \tilde{\nu} \equiv i\nu$ we obtain the following geometry:
\be
ds^2 = \frac{\ell^2}{(3-\tilde{\nu}^2)}\left[ d\rho^2 + \cosh^2\rho d\tau + \frac{4\tilde{\nu}^2}{(3-\tilde{\nu}^2)} \left( du - \sinh \rho d\tau \right)^2 \right]
\ee
which can be interpreted as a squashed three-dimensional hyperbolic geometry. We can, as usual, also consider global identifications of the above metric along Killing directions. This geometry has negative curvature for all values of real $\tilde{\nu}$ and is a solution to the following Euclidean equations of motion
\be
%\left(R_{ij} - \frac{1}{2}R g_{ij} - \frac{1}{\ell^2}g_{ij}\right) + \frac{i}{\mu } C_{ij} = \mathcal{G}_{ij}
\left(R_{ij} - \frac{1}{2}R g_{ij} - \frac{1}{\ell^2}g_{ij}\right)  + \frac{1}{\tilde{\mu}} C_{ij} = 0
\ee
Note that these equations of motion with a negative cosmological constant are \emph{different} than those discussed recently in the context of Chiral Gravity \cite{Maloney:2009ck} which are complex.

\section{Scalar Waves in $WAdS_3$}

In this section we wish to analyze asymptotic solutions to the massive wave equations for a scalar in the background of spacelike $WAdS_3$. Our reason for studying massive scalars is that pure TMG has a single propagating degree of freedom \cite{Deser:1991qk,Park:2008yy,Carlip:2008qh,Grumiller:2008pr,Blagojevic:2008bn} which at the linearized level behaves like a `massive scalar' in the background geometry. The linearized massive wave equation we wish to consider is given by\footnote{Scalar wave equations in the warped black hole background \ref{eq:ng} have been considered in \cite{Oh:2009if,Chakrabarti:2009ww,Oh:2008tc}.}
\be
g^{\mu \nu} \nabla_\mu \nabla_\nu \phi = m^2 \phi\label{3d eoms}
\ee

Choosing an ansatz of the form $\phi(t,r,u) = e^{-i \omega t + i k u} y(r)$ with $k\in\mathbb{R}$ the solution of \ref{3d eoms} is given by a linear combination of two hypergeometric functions\footnote{This solution can be found in \cite{Kim:2008xv} for a scalar in $WAdS_3$ and $WdS_3$ since we can interpret these scalars as charged massive scalars in $AdS_2$ and $dS_2$ as we shall see. We present the explicit solutions, along with a brief discussion of scalars in timelike $WAdS_3$ in appendix \ref{appexpl}.}. Asymptotically we find that the $WAdS_3$ solution behaves as
\be
y(r) \sim |r|^{-\lambda_{\pm}}; \quad \lambda_\pm = {\frac{1}{2} \pm \frac{\sqrt{1 + 4 m^2 L^2 + 4 k^2(1-a^2)/a^2}}{2}}\label{cw}
\ee
where
\be
a^2 = 4\nu^2/(\nu^2+3\epsilon); \quad L^2 =  \ell^2/(\nu^2+3\epsilon)
\ee
where we have defined $\epsilon \in \{ 0,\pm1 \}$ such that $\Lambda = -\epsilon/\ell^2$. Notice that when $a^2=1$, the falloff becomes $k$ independent and matches that of $AdS_3$. Indeed, for $a^2=1$ and $u \sim u+2\pi$ the explicit solutions to \ref{3d eoms} (given in the appendix for all values of $a^2$) have been analyzed in \cite{Balasubramanian:2003kq}. The solutions take the form of a sum of two independent hypergeometric functions with the above falloffs. Requiring that the scalar be single valued in the complex coordinate plane leads to a quantization condition for the normalizable modes of the form
\be
\omega_n = \pm (\lambda_+ + n); \quad n \in \mathbb{Z}^+ \cup \{ 0 \}\label{quant}
\ee
If this condition is imposed, \cite{Balasubramanian:2003kq} concluded that the normalizable solutions become precisely descendants of the highest weight modes which we discuss in \ref{hwmodes}. We find an identical condition for scalars in $WAdS_3$, as discussed in appendix \ref{appexpl}.

When $a^2 > 1$ the falloff becomes complex for large values of $k$ and when $a^2 < 1$ the falloff remains real for all $m^2 L^2 > -1/4$. A complex $\lambda_\pm$ implies that there is an ingoing or outgoing flux at the boundary. We will address this issue in section \ref{sura}.

\subsection{Highest Weight Solutions and Conformal Weights}\label{hwmodes}

Given the isometries of the background solution and in light of the quantization condition \ref{quant}, we are lead to consider solutions in the highest weight representation of the $SL(2,\mathbb{R})$. To do so, we impose the first order condition
\be
\tilde{L}_+ |\phi \rangle = 0, \quad \tilde{L}_0 | \phi \rangle = h_{sc} | \phi \rangle
\ee
where $\tilde{L}_{\pm}$ and $\tilde{L}_0$ are given in the appendix and $h_{sc}$ is the conformal weight. The wave equation can then be written as
\be
\frac{1}{2}\left(\tilde{L}_+ \tilde{L}_- + \tilde{L}_- \tilde{L}_+\right) - \tilde{L}^2_0 - \frac{(1-a^2)}{a^2}\partial^2_u = -{m^2 L^2}
\ee
Once again, choosing an $e^{i k u}$ dependence we find the expected dispersion relation
\be
h_{sc} = {\frac{1}{2} \pm \frac{\sqrt{1 + 4 m^2 L^2 + 4 k^2(1-a^2)/a^2}}{2}}\label{cw}
\ee
The explicit form of the solution is given by
\be
\phi(\tau,r,u) = A_{k h_{sc}} e^{-i h_{sc} \tau + i k u} e^{k \tan^{-1} r} (1+r^2)^{-h_{sc}/2}
\ee
The descendant modes are obtained by applying the lowering operators $\tilde{L}_{-}$ to the above highest weight modes. Their energy is given by $\omega_n = h_{sc} + n$ where $n$ is a positive integer, and they share the same large $r$ asymptotics. Thus the highest weight modes naturally satisfy the condition \ref{quant}.

Note that if we require the energy, $h_{sc}$, to be real then we require $\lambda_\pm$ to be real which boils down to the condition
\be
m^2 L^2 + k^2(1-a^2)/a^2 > -1/4\label{bfw}
\ee
For $a^2=1$ this is simply the Breitenlohner-Freedman bound for three-spacetime dimensions as we should expect. For $a^2 < 1$ the bound is again always satisfied in the region $m^2 L^2 > -1/4$. For $a^2 > 1$ however the bound is not satisfied for large values of $k$ and this is qualitatively different from the case of $AdS_3$.

\subsection{Normalizability}

Given the scalar wave equation, we can define a conserved inner product given by
\be
(u_1,u_2) = i \int_\Sigma d^2 x\sqrt{-g} g^{\tau \mu} (u_1^* \partial_\mu u_2 - u_2 \partial_\mu u_1^*)
\ee
where $u_1$ and $u_2$ are solutions to the wave equation and $\Sigma$ is a fixed $\tau$ slice. If our mode is of the form $\phi = e^{-i \omega \tau + i k u}f(r)$ then the norm is given by
\be
(\phi,\phi) = 2 a L \int_\Sigma d^2 x \left(\frac{(k r + \omega) |f(r)|^2}{1+r^2} \right)
\ee
For $k \neq 0$ and large values of $r$ we have
\be
(\phi,\phi) \sim \int d^2x \frac{k}{r^{1+(h+h^*)}}
\ee
and thus our modes are normalizable for $\Re{[h]} > 0$ provided that we are considering wave packets with finite support in the $u$-direction. Thus for small enough values of $k$ and positive $m^2$, the $\lambda_-$ modes are not normalizable whereas the $\lambda_+$ modes are always normalizable. In the region $-1/4 < m^2 L^2 < 0$ both sets of modes are normalizable.

\subsection{Massive Gravitons}

In \cite{Anninos:2009zi} the conformal weight of the massive gravitons was obtained
\begin{eqnarray}
h_{grav} = {\frac{1}{2} \pm \frac{\sqrt{1 + 4(1-a^2) + 4 k^2(1-a^2)/a^2}}{2}}
\end{eqnarray}
The above conformal weight has precisely the form of the conformal weight of the scalar \ref{cw} so long as the graviton has the following mass
\be
m_g^2 L^2 = (1-a^2)
\ee
This resonates well with the notion that the massive gravitons in TMG are given by a single scalar degree of freedom. Indeed, it was found in \cite{Kim:2009xx} that for $k=0$ the metric perturbation obeys a scalar wave equation with the above mass. What is interesting is that the mass of graviton becomes tachyonic for $a^2 > 1$ and saturates the $k=0$ Breitenloner-Freedman bound at $a^2 = 5/4$.

\subsection{Superradiance}\label{sura}

We can delve somewhat further into the notion of a complex $\lambda_\pm$ by reproducing an argument in \cite{Bardeen:1999px}. In the regime of complex $\lambda_\pm$, we can use the WKB approximation to obtain the effective wavenumber $\kappa$. We can read this off from the wave equation
\be
\kappa = -\frac{i}{y(r)}\frac{d y(r)}{dr} = \pm \frac{1}{(1+r^2)^{1/2}}\left[ \frac{(\omega + k r)^2}{1+r^2}-\frac{k^2}{a^2} \right]^{1/2}\label{kappa}
\ee
assuming masslessness for simplicity. Then the group velocity is given by
\be
v_{group} = \pm \frac{(1+r^2)^{3/2}}{\omega+k r}\left[ \frac{(\omega + k r)^2}{1+r^2}-\frac{k^2}{a^2} \right]^{1/2}
\ee
Notice that $\kappa$ is only real for large $r$ when $a^2 > 1$ and this is the case we should consider. The phase velocity is given by $v_{phase} = \omega/\kappa$ as usual. What \cite{Bardeen:1999px} point out is that for $k < 0$ ($k > 0$) and large positive (negative) $r$ the group and phase velocities have opposite signs.

Consider sending a wave with $k>0$ and $\omega>0$ which is emitted at large positive $r$ and travels in the negative $r$ direction. When $r$ reaches a value where the root in \ref{kappa} changes sign, the wave will encounter a potential barrier. The resultant transmitted wave will have positive group velocity at large negative $r$ which implies a negative phase velocity. Now the energy transport rate carries the same sign as $v_{phase}$ \cite{Bardeen:1999px} implying that there must be energy transported toward $r = \infty$ from the transmitted wave leading to energy flux transmitted through $r = \infty$. Thus, there must be a net flux of energy across $r = \infty$ implying the reflected wave has more energy than the initial emitted wave. We refer to this phenomenon as \textit{superradiance}.

\subsection{Schwinger Pair Production in $AdS_2$}\label{spads2}

If we identify the $u$-direction as $u \sim u + 2\pi$, so that $k$ is quantized, we can interpret our spacetime as an $AdS_2$ background with a constant electric field and an infinite tower of massive Kaluza-Klein excitations. Each Fourier mode of the three-dimensional scalar field is then given by a charged two-dimensional scalar field in an $AdS_2$ background. The equation of motion of such a scalar is given by
\be
\hat{g}^{\mu \nu} \left(\hat{\nabla}_\mu - i q A_\mu\right)\left(\hat{\nabla}_\nu - i q A_\nu\right)\varphi = m^2_2 \ell_2^2 \varphi\label{ads2eom}
\ee
where the hatted metric is that of $AdS_2$ with anti-de Sitter length $\ell_2$ and we are working in the limit of no backreaction. In fact, the above equations of motion are identical to \ref{3d eoms} for each Fourier mode so long as we identify
\be
m_2^2 \ell_2^2 = m^2 L^2 + k^2/a^2; \quad  q A_\mu dx^\mu = k r d\tau
\ee
such that the electric field is given by $q E = k$. Now we can borrow the analysis of \cite{Pioline:2005pf} on charged scalar fields in an $AdS_2$ background with constant electric field. We notice that our condition for superradiance in \ref{bfw} can be expressed as
\be
m_2^2 \ell_2^2 < -1/4 + (q E)^2
\ee
which is precisely the condition for Schwinger pair production in $AdS_2$\footnote{A similar story is found for scalars in the NHEK geometry \cite{StromPrep}}. Note that the electric field has to overcome more than just the mass of the charged scalar due to the confining anti-de Sitter potential well \cite{Pioline:2005pf}. The production rate is given by
\be
\Gamma \sim e^{-2\pi\left( q E - \sqrt{(qE)^2 - m_2^2 \ell_2^2 - 1/4}\right)}
\ee

In this picture, the limit $a^2 \to 1$ is the limit in which the contribution of the Kaluza-Klein mass exactly cancels the strength of the electric field and thus the usual Breitenlohner-Freedman bound is obeyed. When $a^2 > 1$ we find that the electric field is outweighs the Kaluza-Klein mass and leads to pair production, whereas when $a^2 < 1$ the Kaluza-Klein mass outweighs the electric field and there is no pair production. The above observations fit in nicely with the idea that for $a^2 > 1$ timelike geodesics are no longer confined and that the graviton mass becomes tachyonic.

\section{Scalar Waves in $WdS_3$}

We now proceed to study scalars in $WdS_3$. Choosing once again an ansatz of the form $\phi(t,r,u) = e^{-i \omega t + i k u} y(r)$ we find asymptotically that the $WdS_3$ solution behaves as
\be
y(r) \sim r^{-\omega_{\pm}}; \quad \omega_\pm = {\frac{1}{2} \pm \frac{\sqrt{1 - 4 m^2 L^2 - 4 k^2(1+a^2)/a^2}}{2}}
\ee
where
\be
a^2 = 4\nu^2/(3-\nu^2), \quad L^2 =  \ell^2/(3-\nu^2) %, \quad \Lambda>0
\ee
Note that $r$ is asymptotically a timelike coordinate and thus the interpretation of the falloffs is rather different. %There is no longer any superradiance like effect. 
What we find are modes that decay in time with an oscillatory behavior, which resembles the behavior of modes in a regular de Sitter background \cite{Strominger:2001pn,Spradlin:2001pw}. The falloffs are real for a small window of parameter space given by
\be
m^2 L^2 + k^2 \frac{(1+a^2)}{a^2} < 1/4\label{wdsbound}
\ee
If we try to obtain highest weight solutions for the above wave equation we find that the conformal weight becomes complex when the \ref{wdsbound} bound in violated. We will refrain from interpreting the complexity of the conformal weights as an indication that the theory is non-unitary or whether we should consider different representations of the isometry group for future work.

\subsection{Schwinger Pair Production in $dS_2$}

Once again, if we identify $u \sim u + 2\pi$ we can interpret $WdS_3$ as $dS_2$ with an electric field and infinite tower of massive Kaluza-Klein modes. Each Fourier mode of our three-dimensional scalar becomes a charged massive two-dimensional scalar. The condition for Schwinger pair production becomes \cite{Garriga:1993fh,Villalba:1995za,Kim:2008xv}
\be
m_2^2 \ell_2^2 + (q E)^2 > 1/4 \label{bfwds}
\ee
where again $m_2^2 \ell^2_2 = m^2 L^2 + k^2/a^2$ and $q E = k$. Of course particle production in $dS_2$ is not a novel phenomenon and it happens even in the absence of an electric field for large enough values of the mass \cite{Candelas:1975du,Lapedes:1977ip,Mottola:1984ar,Allen:1985ux,Bousso:2001mw}. The presence of the electric field simply enhances this effect.

The reason particles are created in the context of $dS_2$ is related to the mismatch between the ingoing and outgoing vacua leading to a non-trivial Bogoliubov transformation. The condition \ref{bfwds} translates to the requirement of a real absolute value squared of the Bogoliubov coefficient \cite{Kim:2008xv}. On the other hand, $AdS_2$, which is a time independent background exhibits no particle production in the absence of an electric field.

\subsection{Schwinger Pair Production in $\mathbb{R}^{1,1}$}

We can also analyze scalar waves in the $W\mathbb{R}^{2,1}$ background \ref{wflat} where we find that the $k^{th}$ mode of a massive three-dimensional scalar reduces to a charge massive scalar in two-dimensional flat space with parameters
\be
m^2_2 \ell^2 = m^2 \ell^2 + {k^2}/{12}; \quad q A_\mu dx^\mu = k r dt
\ee
such that again the electric field is given by $q E = k$. It is interesting to note that in contrast to the case of charged particles in $AdS_2$, Schwinger pair production in $\mathbb{R}^{1,1}$ occurs non-perturbatively for arbitrarily small electric field with a production rate
\be
\Gamma \sim e^{-m_2^2 \ell^2/2 q E}.
\ee

We can solve explicitly for scalars in $W\mathbb{R}^{2,1}$ to find
\be
\phi(t,r,u) = e^{-i\omega t + i k u} \left( D_{\nu_+} (z) + D_{\nu_-} (-z^*) \right)
\ee
where
\be
\nu_\pm \equiv \frac{1}{2} \left(-1\pm\frac{i \ell^2 m^2}{k}\pm\frac{i k}{12}\right), \quad z =  \frac{(1 + i) (k r+\omega )}{\sqrt{k}}
\ee
and $D_\nu(z)$ is the parabolic cylinder function.

\section*{Acknowledgements}
This work was partially funded by a DOE grant DE-FG02-91ER40654. It has been a great pleasure to discuss this work with F. Denef, M. Esole, M. Guica, T. Hartman, M. Padi and W. Song. I would especially like to thank M. Esole for carefully reading the draft. I would also like to thank the Erwin Schrodinger Institute of Mathematical Physics for its kind hospitality while part of this work was completed.

\appendix

\section{Highest Weight Isometries}

The Killing vectors for the $SL(2,\mathbb{R})_R$ of $WAdS_3$ are
\begin{eqnarray}
\tilde{L}_{\pm} &=& \pm e^{\mp i\tau} \left(\frac{r}{\sqrt{1+r^2}} \,\p_{\tau} \pm i \sqrt{1+r^2} \,\p_r + \frac{1}{\sqrt{1+r^2}} \,\p_u \right)\\
\tilde{L}_0 &=&   -i\, \p_{\tau}
\end{eqnarray}
and they obey the usual $SL(2,\mathbb{R})$ algebra under Lie brackets:
\be
[\tilde{L}_+,\tilde{L}_{-}] = 2 \tilde{L}_0, \quad [\tilde{L}_{\pm },\tilde{L}_0]= \pm  \tilde{L}_{\pm }, \quad  [J_2,\tilde{L}_i] =0, \quad i \in \{ 0,\pm  \}
\ee

\section{Coordinate Transformation}\label{appbcwds}

The metrics \ref{wdsbh} and \ref{wdsbc} are related by the coordinate transformation
\be
r = \tilde{r} + \frac{1}{2}(r_+ - r_-)
\ee
and we relate the parameters as
\be
\omega = \frac{2\nu^2}{3(\nu^2+1)} \left( r_+ - r_- + \frac{\sqrt{r_+ r_-(-\nu^2+3)}}{\nu} \right); \quad r_h^2 = \frac{1}{4}(r_+ + r_-)^2
\ee
Recall, that the above parameters are relevant in the region $\nu^2 < 3$. Imposing $r_+ > 0$ and $r_- < 0$ to be real implies that $\omega^2 < 4\nu^2 r_h^2/3(\nu^2+1)$ in the \ref{wdsbh} coordinate system. This is the region that contains CTCs.

\section{Explicit Solutions, Timelike $WAdS_3$ Scalars}\label{appexpl}

\subsection{Explicit Solutions}

We obtain the explicit solutions for the $k^{th}$ Fourier mode of the three-dimensional massive scalars by realizing they are solutions to the following two-dimensional wave equation given in \ref{ads2eom}
\be
\hat{g}^{\mu \nu} \left(\hat{\nabla}_\mu - i q A_\mu\right)\left(\hat{\nabla}_\nu - i q A_\nu\right)\varphi = m^2_2 \ell_2^2 \varphi
\ee
where
\be
m_2^2 \ell_2^2 = m^2 L^2 + k^2/a^2; \quad  q A_\mu dx^\mu = k r d\tau
\ee
The hatted quantities refer to $AdS_2$ and we are assuming $u \sim u + 2\pi$. The solutions to the above equation we obtained in \cite{Kim:2008xv} and are given by
\begin{multline}
\varphi_\omega(\tau,r) = e^{-i\omega \tau} z^{n/2}(1-z)^{n^*/2} \\ \times \left[ c_1 F(\mu,\nu;\gamma;z) + c_2 z^{1-\gamma} F(\mu-\gamma+1,\nu-\gamma+1;2-\gamma;z)  \right]
\end{multline}
with
\begin{eqnarray}
n &=& \frac{1}{2} - {\omega} + i k \\
\mu &=& \frac{n+n^*}{2} - i \sqrt{{k^2} - m_2^2 \ell_2^2 - 1/4}, \quad \nu = \mu^*\\
\gamma &=& n + \frac{1}{2}; \quad z = \frac{1+i r}{2}
\end{eqnarray}

In fact, redoing the analysis of \cite{Balasubramanian:2003kq} for the above solution we find that demanding singled valued solutions in the complex coordinate plane, there is a quantization condition for the normalizable solutions of the form
\be
\omega_n = \pm (\lambda_+ + n); \quad n \in \mathbb{Z}^+ \cup \{ 0 \}
\ee
in the region where $\lambda_+$ is real. This follows from the fact that our scalar equation of motion \ref{3d eoms} is related to (3.11) in \cite{Balasubramanian:2003kq} by replacing their $\mu^2$ with $(4 m^2 L^2 + 4 k^2(1 - a^2)/a^2)$. Thus, following \cite{Balasubramanian:2003kq}, we can relate the above solutions to the highest weight solutions and their descendants obtained in \ref{hwmodes}.

Similarly, \cite{Kim:2008xv} obtained the solution for a charged massive scalar in $dS_2$ and the solution is related to the above solution by an analytic continuation. In this case, we have
\begin{multline}
\varphi_\kappa(\phi,r) = e^{i \kappa \phi} z^{n/2}(1-z)^{n^*/2} \\ \times \left[ c_1 F(\mu,\nu;\gamma;z) + c_2 z^{1-\gamma} F(\mu-\gamma+1,\nu-\gamma+1;2-\gamma;z)  \right]
\end{multline}
with
\begin{eqnarray}
n &=& \frac{1}{2} - {\kappa} + i k\\
\mu &=& \frac{n+n^*}{2} - i \sqrt{{k^2} + m_2^2 \ell_2^2 - 1/4}, \quad \nu = \mu^*\\
\gamma &=& n + \frac{1}{2}; \quad z = \frac{1+i \sinh\tau}{2}
\end{eqnarray}

\subsection{Timelike $WAdS_3$ Scalars}\label{apptime}

The treatment is very similar to the spacelike case. The global timelike $WAdS_3$ metric in global coordinates is given by
\be
ds^2 = L^2 \left[ (1+r^2)du^2 + \frac{dr^2}{1+r^2} - a^2 \left( d\tau + r d u \right)^2 \right]
\ee
with isometries
%\begin{eqnarray}
%L_{\pm} &=&  e^{\pm u} \left(  \frac{1}{\sqrt{r^2+1}}\partial_\tau \mp \sqrt{r^2+1} \partial_r -  \frac{r}{\sqrt{r^2+1}} \partial_u \right)\\
%L_0 &=& \partial_u; \quad  \partial_\tau
%\end{eqnarray}
\begin{eqnarray}
J_0 &=& \frac{2\sinh u}{\sqrt{r^2+1}}\partial_\tau - 2\cosh u\sqrt{r^2+1} \partial_r +  2\tanh u\frac{r}{\sqrt{r^2+1}} \partial_u \\
J_1 &=& -\frac{2\cosh u}{\sqrt{r^2+1}}\partial_\tau + 2\sinh u\sqrt{r^2+1} \partial_r -  2\tanh u\frac{r}{\sqrt{r^2+1}} \partial_u \\
J_2 &=& 2\partial_u
\end{eqnarray}
which obey $[J_1,J_2] = 2 J_0$, $[J_0,J_1]= - 2 J_2$, $[J_0,J_2] = 2 J_1$ and $[J_i,\partial_\tau] = 0$.
%which obey $[L_+,L_{-}] = 2 L_0$, $[L_{\pm },L_0]= \pm  L_{\pm }$ and $[\partial_\tau,L_j] =0$.

Choosing the ansatz $\phi(\tau,r,u) = e^{- i \omega \tau + i k u} f(r)$, we give here the asymptotic behavior of the solutions near $r = \pm \infty $
\be
\phi(\tau,r,u) \sim e^{i k u - i \omega \tau} |r|^{\lambda_\pm}; \quad \lambda_\pm = \frac{1}{2} \pm \frac{\sqrt{1+4m^2L^2+4\omega^2(a^2-1)/a^2}}{2}
\ee
Note that $\lambda_\pm$ now depend on the frequency $\omega$ rather than the $k$ and are always real only for $a^2 > 1$.

\end{document}